\documentclass[preprint,prd,amsmath,amssymb,floatfix,nofootinbib]{revtex4}
\usepackage[colorlinks=true,urlcolor=black]{hyperref}

\hypersetup{backref = true,pagebackref = true,hyperindex = true, colorlinks = true,
  breaklinks = true, urlcolor = blue, linkcolor = blue, bookmarks = true,
  bookmarksopen = true, citecolor=red}

\usepackage{bm}
\usepackage{color}
\usepackage{dcolumn}
\usepackage{graphics}
\usepackage{graphicx}
\usepackage{subfigure}
\usepackage{slashed}
\usepackage{placeins}

\usepackage{pdfpages}
\usepackage{eso-pic}
\usepackage{everyshi}
\usepackage{pdflscape}

\newcommand{\bea}{\begin{eqnarray}}
\newcommand{\eea}{\end{eqnarray}}
\newcommand{\be}{\begin{equation}}
\newcommand{\ee}{\end{equation}}

\newcommand{\ar}{a_s}

\begin{document}

\title{On Bjorken sum rule: heavy quarks and analytic coupling
}
\author{I.R. Gabdrakhmanov$^{1}$, N.A Gramotkov$^{1,2}$, A.V.~Kotikov$^{1}$,  O.V.~Teryaev$^{1}$, D.A. Volkova$^{1,3}$  and I.A.~Zemlyakov$^{4}$}
\affiliation{
  $^1$Bogoliubov Laboratory of Theoretical Physics,
  Joint Institute for Nuclear Research, 141980 Dubna, Russia;\\
$^2$Moscow State University, 119991, Moscow, Russia\\
  $^3$Dubna State University,
  141980 Dubna, Moscow Region, Russia;\\
$^4$Department of Physics, Universidad Tecnica Federico Santa Maria, Avenida Espana 1680, Valparaiso, Chile}


\begin{abstract}

We present the results of \cite{Heavy}, where good agreement was obtained
between  calculations within the framework of analytic QCD and experimental data on polarized
Bjorken sum rule. The heavy quark contributions are taken into account.

\end{abstract}

\maketitle

\section{Introduction}

Experimental data for the polarized Bjorken sum rule (BSR) $\Gamma^{p-n}_1(Q^2)$ \cite{Bjorken:1966jh} were obtained in a wide range
of squares of spacelike momenta $Q^2$: 0.021 GeV$^2\leq Q^2 <$5 GeV$^2$ (see \cite{Deur:2021klh,Gabdrakhmanov:2024bje} and references therein),
which opens up broad possibilities for studying QCD at low $Q^2$ \cite{Deur:2018roz}.

In theory, over the last thirty years, an extension of the QCD coupling constant {\it (couplant)} has been developed that does not have a
Landau singularity for low $Q^2$ and is called analytic perturbation theory (APT) \cite{ShS,BMS1}.
APT has already been used to compare theoretical expressions and experimental BSR data
\cite{Gabdrakhmanov:2024bje,Pasechnik:2008th,Ayala:2017uzx,Ayala:2018ulm,Gabdrakhmanov:2023rjt}.

In this paper we give a brief overview of the application \cite{Heavy} of
heavy quark (HQ) contribution to the BSR calculated
at the two-loop level in Ref. \cite{Blumlein:2016xcy}. This study was carried out in the framework of APT, where the photoproduction
limit was also investigated.

\section{Bjorken sum rule}

The polarized BSR is defined as the difference of the polarized structure functions of the proton and neutron,
integrated over the entire interval $x$
\be
\Gamma_1^{p-n}(Q^2)=\int_0^1 \, dx\, \bigl[g_1^{p}(x,Q^2)-g_1^{n}(x,Q^2)\bigr]
\label{Gpn} 
\ee
and can be represented as:
\be
\Gamma_1^{p-n}(Q^2)=
\frac{g_A}{6} \, \bigl(1-D_{\rm BS}(Q^2)\bigr) +\frac{\hat{\mu}_4 M^2}{Q^{2}+M^2} \, ,
\label{Gpn.mOPE} 
\ee
where $g_A$=1.2762 $\pm$ 0.0005 \cite{PDG20} is the axial charge of the nucleon, $(1-D_{BS}(Q^2))$ is the contribution of the leading twist (twist-2),
and $(\hat{\mu}_4 M^2)/(Q^{2}+M^2)$ is the so-called "massive" representation for twist-four (see \cite{Teryaev:2013qba}).

According to \cite{Cvetic:2006mk}  (in $k$-order of pertturbation theory (PT)) we introduce and use here derivatives 
\be
\tilde{a}^{(k)}_{n+1}(Q^2)=\frac{(-1)^n}{n!} \, \frac{d^n a^{(k)}_s(Q^2)}{(dL)^n},~~a^{(k)}_s(Q^2)=\frac{\beta_0 \alpha^{(k)}_s(Q^2)}{4\pi}=\beta_0\,\overline{a}^{(k)}_s(Q^2),
\label{tan+1}
\ee
which play an important role when using analytic QCD.
Here and below $\beta_0$ is the first coefficient of the $\beta$-function of QCD
$\beta(\overline{a}^{(k)}_s)=-{\left(\overline{a}^{(k)}_s\right)}^2 \bigl(\beta_0 + \sum_{i=1}^k \beta_i {\left(\overline{a}^{(k)}_s\right)}^i\bigr)$,
where $\beta_i$ are known up to $k=4$ \cite{Baikov:2008jh}.

The series of derivatives $\tilde{a}_{n}(Q^2)$ can be used instead of the series of $\ar$-powers. Indeed, each derivative reduces the $\ar$ power,
but, on the other hand, it produces an additional $\beta$-function proportional to $\ar^2$.
By definition (\ref{tan+1}), in the leading order (LO) the expressions for $\tilde{a}_{n}(Q^2)$ and $\ar^{n}$ exactly coincide.
Beyond LO, a one-to-one correspondence between $\tilde{a}_{n}(Q^2)$ and $\ar^{n}$ was constructed \cite{Cvetic:2006mk,Cvetic:2010di}
and extended to the fractional case in \cite{GCAK}.

The perturbative expansion up to the $k$th order has the following form
\be
D^{(1)}_{\rm BS}(Q^2)=\frac{4}{\beta_0} \, \tilde{a}^{(1)}_1,~~D^{(k\geq2)}_{\rm BS}(Q^2)=
\frac{4}{\beta_0} \, \left(\tilde{a}^{(k)}_{1}+\sum_{m=2}^k\tilde{d}_{m-1}\tilde{a}^{(k)}_{m}
\right),
\label{DBS.1} 
\ee
where $\tilde{d}_1$, $\tilde{d}_2$, and $\tilde{d}_3$ are known from direct calculations (see, e.g., \cite{Chen:2006tw}).
The coefficient $\tilde{d}_4$ is not known exactly, but its value  has been estimated at \cite{Ayala:2022mgz}.

From here on we will take $f=3$. Thus,
\footnote{
  The coefficients $\beta_i$ $(i\geq 0)$ of the QCD $\beta$-function and, as a consequence, the couplant $\alpha_s(Q^2)$
  depend on the number $f$ of active flavors, and each new quark is switched on at the threshold level $Q^2_f$ in accordance with
  \cite{Chetyrkin:2005ia}. The corresponding QCD parameters $\Lambda^{(f)}$ in N$^i$LO PT are obtained in \cite{Chen:2021tjz}.}
\bea
\tilde{d}_1=1.59,~~\tilde{d}_2=2.73,
~~\tilde{d}_3=8.61,~~\tilde{d}_4=21.52 \, .
\label{td123} 
\eea

Switching from conventional PT to APT given by replacement $D^{(1)}_{\rm BS}(Q^2)$ with $\tilde{a}^{(k)}_{m}$ by $D^{(1)}_{\rm A,BS}(Q^2)$ with analytic couplants
$\tilde{A}^{(k)}_{m}$ (the corresponding expressions for $\tilde{A}^{(k)}_{m}$ can be found \cite{Kotikov:2022sos}). So, we have in APT
\be
\Gamma_{\rm{A},1}^{p-n}(Q^2)=
\frac{g_A}{6} \, \bigl(1-D_{\rm{A,BS}}(Q^2)\bigr) +\frac{\hat{\mu}_{\rm{A},4}M^2}{Q^{2}+M^2}.
\label{Gpn.MA} 
\ee

{\bf 3.} {\it HQ contribution}
was calculated in \cite{Blumlein:2016xcy} only at the next-to-leading (NLO) order, that
leads to the following replacement for $\tilde{d}_1$:
\be
\tilde{d}_1 \to \tilde{d}_1 \textcolor{red}{-} \sum_{i=c,b,t}\,C_1(\xi_i)\,,~~\xi_i=\frac{Q^2}{m^2_i}~~(i=c,b,t)
    \label{Gpn.MA.HQ} 
\ee
and $m_c=1.27$ GeV, $m_b=4.18$ GeV and $m_t=172.76$ GeV (see \cite{PDG20}).

$C_1(\xi)$ has the following form
\bea
&&C_1(\xi)=
\frac{8}{3\beta_0}
\, \biggl\{\frac{6\xi^2+2735\xi+11724}{5040\xi}- \frac{3\xi^3+106\xi^2+1054\xi+4812}{2520\xi}\,L(\xi)
-\frac{5}{3\xi(\xi+4)}\,L^2(\xi)\nonumber\\ 
&&+\frac{3\xi^2+112\xi+1260}{5040}\,\ln(\xi)\biggl\},~
L(\xi)=\frac{1}{2\delta}\,\ln\left(\frac{1+\delta}{1-\delta}\right),~\delta^2=\frac{\xi}{4+\xi}\,.
\label{C1} 
\eea

At large $Q^2$ values, $C_1(\xi) \sim 1/(3\beta_0)$ and at low $Q^2$ values, $C_1(\xi) \sim 2/(3\beta_0)\ln(\xi)$, i.e. it
rises as $\ln Q^2$ at $Q^2 \to 0$.

\section{Results}

\begin{table}[h!]
\begin{center}
\begin{tabular}{|c|c|c|c|}
\hline
& $M^2$ for (\ref{Gpn.MA})& $\hat{\mu}_{\rm{MA},4}$& $\chi^2/({\rm d.o.f.})$  for \\
& (for (\ref{Gpn.MAn})) &for (\ref{Gpn.MA}) &  (\ref{Gpn.MA}) (for (\ref{Gpn.MAn})) \\
 \hline
 LO & 1.631 $\pm$ 0.301 (0.576 $\pm$ 0.046) & -0.166 $\pm$ 0.001 & 0.789 (0.575) \\
 \hline
 NLO & 1.740 $\pm$ 0.389  (0.411 $\pm$ 0.035) & -0.143 $\pm$ 0.002 & 0.742 (0.630) \\
 \hline
 N$^2$LO & 1.574 $\pm$ 0.319 (0.400 $\pm$ 0.034) &-0.144 $\pm$ 0.002 & 0.714 (0.621) \\
 \hline
 N$^3$LO & 1.587 $\pm$ 0.327  (0.411 $\pm$ 0.035) & -0.145 $\pm$ 0.002 & 0.733 (0.618) \\
   \hline
N$^4$LO & 1.630 $\pm$ 0.344  (0.412 $\pm$ 0.035) &-0.144 $\pm$ 0.002 & 0.739  (0.621) \\
 \hline
\end{tabular}
\end{center}
\caption{%
Fitting parameters.
}
\label{Tab:BSR1}
\end{table}

Since the usual PT is not applicable for BSR with small $Q^2$
(see \cite{Gabdrakhmanov:2024bje,Pasechnik:2008th,Ayala:2017uzx,Ayala:2018ulm,Gabdrakhmanov:2023rjt}),
we consider here only APT. Moreover, here we restrict ourselves to analyses for small $Q^2$. A more general case can be found in our
paper \cite{Heavy}.
The results of fitting experimental data obtained only with statistical uncertainties are presented in Table \ref{Tab:BSR1} and shown in Fig. 1.

Our results obtained for different APT orders are almost equivalent: the corresponding curves are indistinguishable
when $Q^2\to 0$, and differ slightly for other values of $Q^2$. As can be seen from Fig. 1, the quality of the fit is quite good,
which is also demonstrated by the values of the corresponding $\chi^2/({\rm d.o.f.})$ (see Table \ref{Tab:BSR1}).

As in the case without heavy quarks, considered earlier in \cite{Gabdrakhmanov:2024bje,Gabdrakhmanov:2023rjt},
the results are not entirely satisfactory (see Fig. 1).
The curves obtained as a result of the fitting take negative values when we go to very low values of $Q^2$: $Q^2 <$0.02 GeV$^2$.


\begin{figure}[t]
\centering
\includegraphics[width=0.98\textwidth]{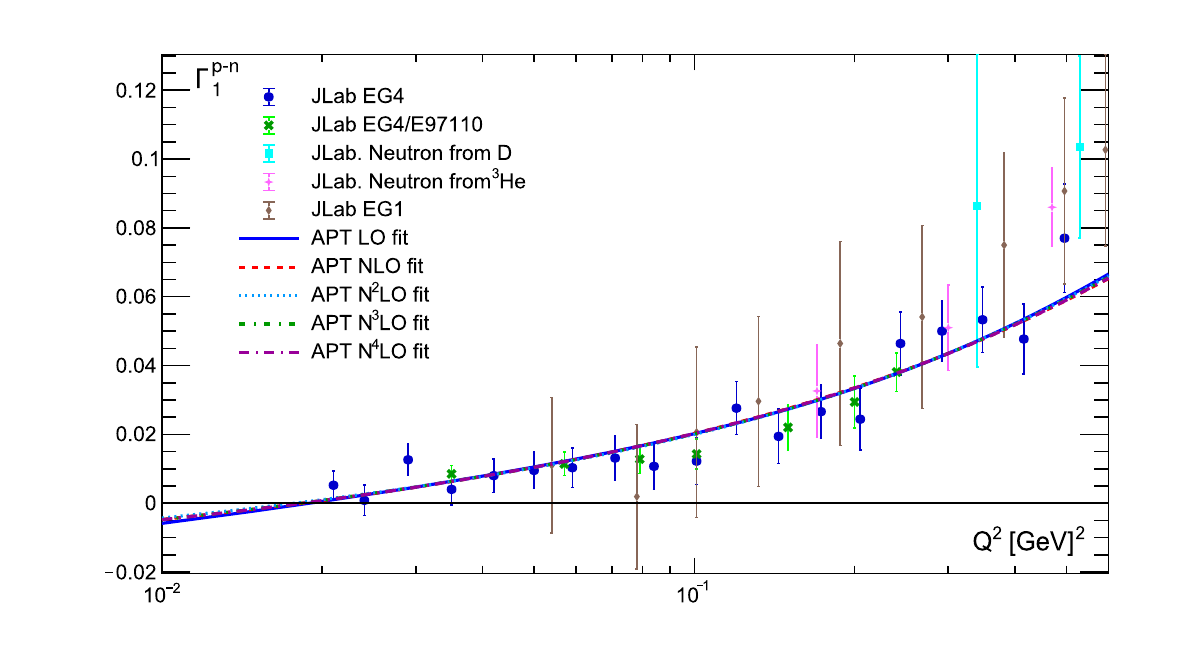}
\caption{
  \label{fig:low}
  The results for $\Gamma_1^{p-n}(Q^2)$ (\ref{Gpn.MA}) in the first five orders of APT 
from fits of experimental data
with $Q^2 <$0.6 GeV$^2$.
}
\end{figure}


{\it Photoproduction.}
To obtain the
limit of $\Gamma_{\rm{A},1}^{p-n}(Q^2\to 0)\to 0$ following from the finiteness of the photoproduction cross section, a new form for $\Gamma_{\rm{A},1}^{p-n}(Q^2)$ was proposed in \cite{Gabdrakhmanov:2024bje}:
\be
\Gamma_{\rm{A},1}^{p-n}(Q^2)=
\, \frac{g_{A}}{6} \, \bigl(1-D_{\rm{A,BS}}(Q^2) \cdot \frac{Q^2}{Q^2+M^2}\bigr) +
\frac{\hat{\mu}_{\rm{A},4}M^2}{Q^{2}+M^2}+\frac{\hat{\mu}_{\rm{A},6}M^4}{(Q^{2}+M^2)^2},~~
\label{Gpn.MAn} 
\ee
where we added the ``massive'' twist-six term and introduced a factor $Q^2/(Q^2+M^2)$ to modify the twist-two contribution.

The finiteness of the cross section in the limit of a real photon now leads to (see \cite{Heavy})
\be
\hat{\mu}_{\rm{A},6} =-G\,M^2+\frac{5g_{A}}{54},~~
\hat{\mu}_{\rm{A},4} = -\frac{g_{A}}{6} -\hat{\mu}_{\rm{A},6}= G\,M^2-\frac{7g_{A}}{27}\,,
\label{Gpn.MAnQ0.4} 
\ee
where  \cite{Soffer:1992ck} $G=0.0631$
is small and, thus, $\hat{\mu}_{\rm{A},4}<0$ and $\hat{\mu}_{\rm{A},6}>0$.

The results of fitting the theoretical predictions based on (\ref{Gpn.MAn}) with $\hat{\mu}_{\rm{MA},4}$ and $\hat{\mu}_{\rm{MA},6}$
from (\ref{Gpn.MAnQ0.4}) are presented in Table \ref{Tab:BSR1} and Fig. 2.
As can be seen from Table \ref{Tab:BSR1}, the results are very similar to those obtained earlier in
\cite{Gabdrakhmanov:2024bje,Gabdrakhmanov:2023rjt}, since the increase in $C_1(\xi)$ as $Q^2 \to 0$ is compensated
by the decrease in $\tilde{A}^{(k)}_{\nu=2}(Q^2 \to 0)$.

We also see the similarity between the results shown in Fig. 1 and 2. The difference appears only for small values of $Q^2$.

\begin{figure}[t]
\centering
\includegraphics[width=0.98\textwidth]{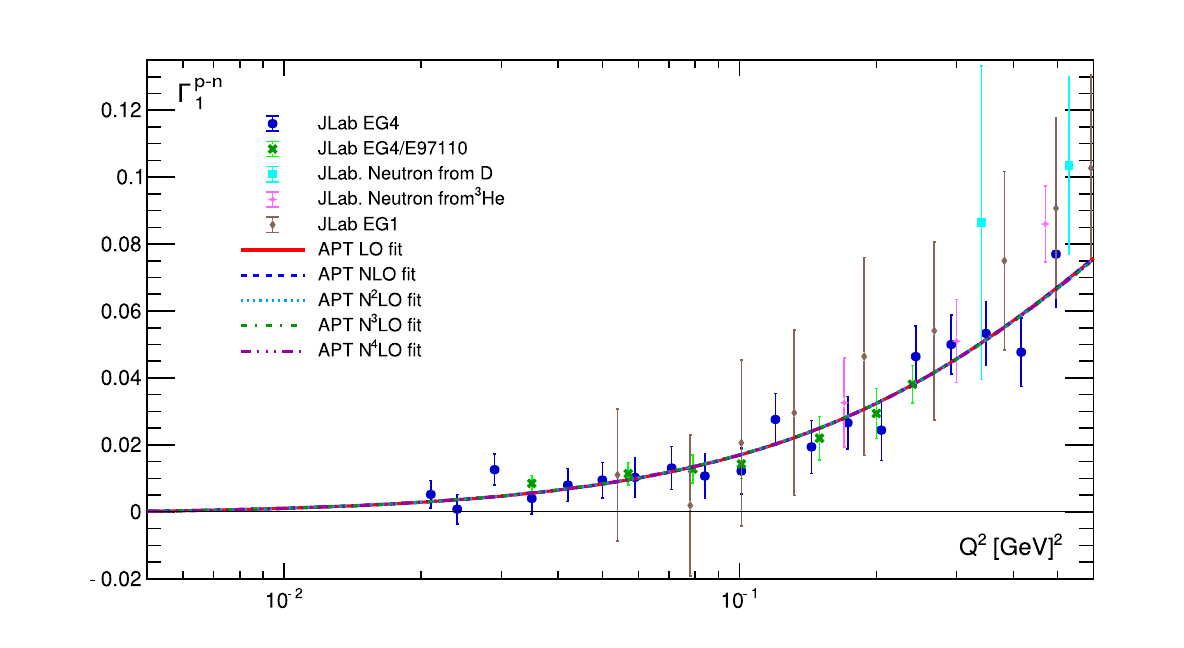}
\caption{
  \label{fig:low1}
  As in Fig. 1 but for (\ref{Gpn.MAn}).
}
\end{figure}

\section{Conclusions}

We gave a brief overview of the results \cite{Heavy}, where we considered 
the Bjorken sum rule $\Gamma_{1}^{p-n}(Q^2)$ in the APT framework in the first five PT orders with the
added contribution of heavy quarks and obtained results similar to those obtained in previous studies
\cite{Pasechnik:2008th,Ayala:2017uzx,Gabdrakhmanov:2023rjt}.

By investigating the low behavior of $Q^2$, we found, as in previous studies without heavy quarks, that there
is a discrepancy between the results obtained in the fits and photoproduction. Indeed, the results of the fits
extended to low $Q^2$ lead to negative values for the Bjorken sum rule:
$\Gamma_{\rm{A},1}^{p-n}(Q^2\to 0) <0$, which contradicts the finiteness of the cross section in the real photon limit,
which corresponds to $\Gamma_{\rm{A},1}^{p-n}(Q^2\to 0) =0$.

To solve the problem, we used a modification (\ref{Gpn.MAn}) of the OPE formula for $\Gamma_{\rm{A},1}^{p-n}(Q^2)$ proposed in
\cite{Gabdrakhmanov:2024bje}.
Using it, we found good agreement between the results obtained in the fits and photoproduction.

{\bf Acknowledgments}~~Authors are grateful to Alexandre P. Deur for
information about new experimental data in Ref. \cite{Deur:2021klh} and and Johannes Blumlein for
initiating the consideration of the contribution of heavy quarks.
This work was supported in part by the Foundation for the Advancement of Theoretical
Physics and Mathematics “BASIS”.
One of us (I.A.Z.)
is supported by the Directorate of
Postgraduate Studies of the Technical University of Federico Santa Maria.




\end{document}